# Copy Propagation subsumes Constant Propagation

*Sreekala S. and Vineeth Paleri*
*National Institute of Technology, Calicut, India*

**Abstract**

Constant propagation and copy propagation are code transformations that may avoid some load operations and can enable other optimizations. In literature, constant and copy propagations are considered two independent transformations requiring two different data flow analyses. Here we give a generic definition for copy propagation which enables us to view constant propagation as a particular case of copy propagation and formulate a novel data flow analysis that unifies these two transformations.

## 1. Introduction

Constant Propagation [2, 3, 4] and Copy Propagation [1, 2, 4] are code transformations that avoid some of the load operations. These transformations, in general, enable other optimizations such as constant folding, dead-code elimination, and common sub-expression elimination. Copy propagation is based on copy statements of the form *x=y* and constant propagation is based on statements of the form *x=c*, where *c* is a constant. The idea of constant propagation is to replace the use of variable *x* with a constant *c*, wherever possible after the statement *x=c*. The idea of copy propagation is similar to constant propagation in which the use of variable *x* after a copy statement *x=y* is replaced with the variable *y*, wherever possible. Now, we can see that constant propagation is a special case of copy propagation where the source operand of the copy statement is a constant instead of a variable. The similarity between constant propagation and copy propagation is not readily visible since the definition of copy propagation in literature is not complete; i.e. current definitions of copy propagation [1, 2, 4] consider only a subclass of possible copy propagations where only a single definition reaches the use of a variable. Here we provide a generic definition for copy propagation which enables the formulation of a new static data flow analysis that unifies constant propagation and copy propagation.

Section 2 of the paper gives the notations used and the definitions required. Section 3 gives the data flow analysis required to collect the necessary information. Section 4 describes the transformation using the information collected by the data flow analysis. Section 5 gives an example, and section 6 concludes the paper.

## 2. Notations and Definitions

### 2.1 Program Representation

The input to the data flow analysis algorithm is a program represented as a Control Flow Graph (CFG), with a unique entry basic block, denoted *entry*, and a unique exit basic block, denoted *exit*. We assume that each basic block contains a single statement in *three address code* form, and the *entry* and *exit* blocks contain empty statements. Here we use *s* to represent the statement in a basic block *B*.

### 2.2 Definitions

**Definition 1[Copy Statement]:** A statement *x=e* is said to be a copy statement if *e* is a constant or a variable.

In the proposed method for data flow analysis, we represent a copy statement $x=y$ as a pair $(x, y)$.

**Definition 2[Constant Propagation]:** The use of a variable $y$ in the statement $z=x+y$ occurring at a point $p$ can be replaced by a constant $c$ if every path from the *entry* node to point $p$ contains the same definition $y=c$, for the variable $y$, and after the definition prior to reaching $p$, there is no redefinition to the variable $y$.

The definitions of copy propagation available the literature [1, 2, 4] are not *complete* in the sense that they consider a subclass of possible copy propagations; i.e. they consider the case where only a single definition reaches the use of a variable.

Let us consider the definition of copy propagation given in [2]:

*"Suppose we have a statement d: t ← z and another statement n that uses t, such as n: y = t $\oplus$ x. If d reaches n, and no other definition of t reaches n, and there is no definition of z on any path from d to n (including a path that goes through n one or more times), then we can rewrite n as n: y ← z $\oplus$ x."*

Note that the above definition considers only the variable definitions that dominate the *use* point of the variable. Here, we give a generic definition for copy propagation, which aligns with the definition of constant propagation available in the literature.

**Definition 3[Copy Propagation]:** The use of a variable $y$ in the statement $z=x+y$ occurring at a point $p$ can be replaced by a variable $w$ if every path from the *entry* node to point $p$ contains the same definition $y=w$, for the variable $y$, and after the definition prior to reaching $p$, there is no redefinition to the variable $y$ and no redefinition to the variable $w$.

**Definition 4[Available Copy Statements]:** A copy statement $x=e$, where $e$ is a constant $c$ or a variable $y$, is said to be *available* at a point $p$ if every path from the *entry* node to point $p$ contains the copy statement $x=e$, and after the copy statement prior to reaching $p$, there is no modification to the variable $x$ and no modification to the variable $y$.

## 3. Data Flow Analysis: Available Copy Statements

Here we formulate a new data flow analysis called *Available Copy Statements* to combine Constant Propagation and Copy Propagation. Available copy statements at each point can be computed statically using a forward data flow analysis. Since we are doing static analysis, the information obtained is conservative.

The information obtained from available copy statements analysis is a set of pairs at each point in the CFG. Each pair of the form $(x, e)$ represents a definition of $x$ with $e$ as a variable or a constant.

For each statement of the form *s: x=e*, first remove all the pairs of the form *(x,\*)* and *(\*, x)* from the information available at the input point of the statement *s*. Then *generate* and add the pair *(x,e)* to the information if $e$ is a constant or a variable. At a confluence point, take the *intersection* of the pairs available at the output points of the predecessors of the statement (see Figure 1(a)).

Using the definition given in the literature [1, 2, 4], we cannot capture the copy propagation in the example shown in Figure 1(a) because they handle only the cases where a single definition reaches a point. But the information collected at the confluence point in Figure 1(a) using the proposed method shows that the copy statement *(y,x)* is *available* at the input point of $s_3$, and copy propagation can be performed as indicated in Figure 1(b).

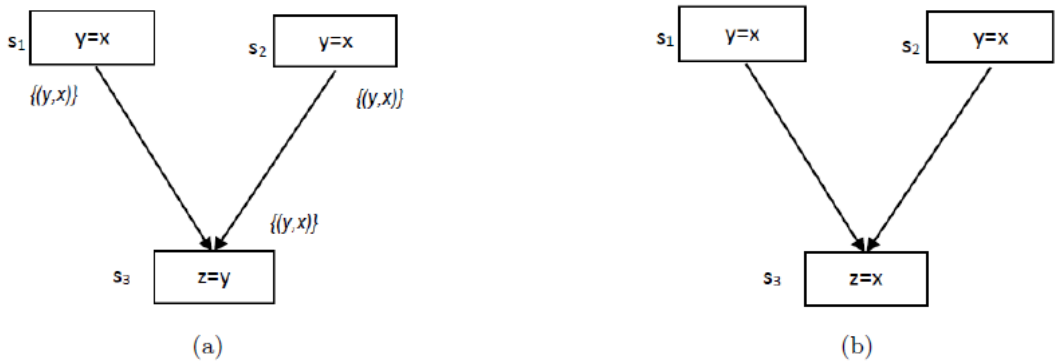

Figure 1: (a) Input CFG annotated with Available Copy Statements (b) Transformed CFG

## 4. Transformation

The basic idea of transformation is that, for each use of a variable *x* in a statement, check whether there exists a pair of the form *(x,\*)* at the input point of the corresponding statement. If a pair of the form *(x,c)*, where *c* is a constant, is available, then *x* can be replaced with *c*. Otherwise, if a pair of the form *(x y)* is available, then check whether there exists a pair of the form *(y,\*)*. The check will be repeated until a constant is obtained or no further checks are possible.

In the literature [1, 2, 4], copy propagation is done by replacing the use of a variable *y* in a statement *s: z=x+y* with its immediate definition available at the input point of *s*. Using this approach, the copy propagation should be repeated multiple times to replace a chain of copy statements (refer Figure 2). But in the proposed method, we do a recursive search to get the best possible replacement for a variable. Therefore, a chain of copy statements can be replaced in a single step, which makes the transformation efficient.

## 5. An Example

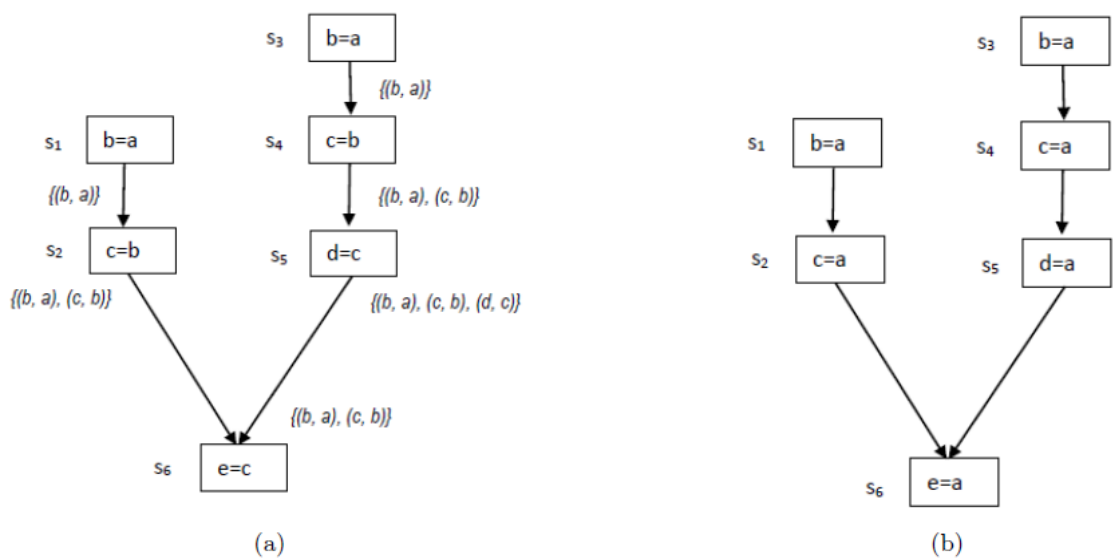

Figure 2: (a) Input CFG annotated with Available Copy Statements (b) Transformed CFG

Consider the copy statement $s_6$: *e=c* in the example shown in Figure 2(a). There is a pair *(c,b)* at the input point. Since *b* is a variable, it searches recursively and gets the variable *a* from the pair *(b,a)*. In the next step, since there is no pair of the form *(a,\*)*, it returns the variable *a* itself. Then, the variable *c* in $s_6$ is replaced with the variable *a*. Similarly, transformations are applied for the copy statements $s_2$, $s_4$, and $s_5$.

## 6 Conclusion

We have given a new definition for copy propagation which is generic unlike the restricted definitions available in the literature. Using this new definition for copy propagation we have formulated a new data flow analysis called *available copy statements* for combining constant propagation and copy propagation. The method is based on the concept of the *availability* of copy statements. The data flow analysis presented here performs more copy propagations than the traditional approach. Also, the transformation is efficient since it performs multiple transformations in a single pass.